\documentclass[twoside]{article}
\usepackage{graphicx}
\usepackage{fancyhdr}
\usepackage{multicol}
\usepackage{styl-ap}

\begin{document}
\title{Expected Hard X-Ray and Soft Gamma-Ray from Supernovae}
\author{Keiichi Maeda\work{1,2}, YukikatsuTerada\work{3}, Aya Bamba\work{4}}
\workplace{Department of Astronomy, Kyoto University, Japan 
\next
Kavli Institute for the Physics and Mathematics of the Universe (WPI), University of Tokyo, Japan
\next
Department of Physics, Saitama University, Japan
\next
Department of Physics and Mathematics, College of Science and Engineering, Aoyama Gakuin University, Japan}
\mainauthor{keiichi.maeda@kusastro.kyoto-u.ac.jp}
\maketitle

\begin{abstract}%
High energy emissions from supernovae (SNe), originated from newly formed radioactive species, provide direct evidence of nucleosynthesis at SN explosions. However, observational difficulties in the MeV range have so far allowed the signal detected only from the extremely nearby core-collapse SN 1987A. No solid detection has been reported for thermonuclear SNe Ia, despite the importance of the direct confirmation of the formation of $^{56}$Ni, which is believed to be a key ingredient in their nature as distance indicators. In this paper, we show that the new generation hard X-ray and soft $\gamma$-ray instruments, on board {\em Astro-H} and {\em NuStar}, are capable of detecting the signal, at least at a pace of once in a few years, opening up this new window for studying SN explosion and nucleosynthesis. 
\end{abstract}

\keywords{Nuclear reaction, nucleosynthesis, abundances - Supernovae: general}

\begin{multicols}{2}
\section{Introduction}
Supernova (SN) explosions trigger (or are triggered by) explosive nucleosynthesis, and they are believed to be main production sites of heavy elements in the Universe. The resulting yields are sensitive to explosion mechanism(s), and thus studying nucleosynthesis products is important to uncover the still-debated explosion mechanism. Especially important is the production of $^{56}$Ni -- this is the origin of Fe (as a result of the radioactive decay chain $^{56}$Ni $\to$ $^{56}$Co $\to$ $^{56}$Fe), and the decay is believed to provide a source of emissions from (many classes of) SNe through thermalization of emitted $\gamma$-rays and positrons. In type Ia supernovae (SNe Ia), about half of an exploding white dwarf in mass is processed into $^{56}$Ni, supporting their huge luminosities as distance indicators. However, the most direct evidence in this scenario is still missing -- there has been no solid detection of the decay $\gamma$-rays from SNe except for SN 1987A (e.g., Dotani et al., 1987; Sunyaev et al., 1987). Especially, no solid detection has been reported for SNe Ia (see Milne et al., 2004 for a review). 

From a theoretical point of view, studying this high energy emission has been restricted to one-dimensional models (see Milne et al., 2004, for a review) despite the importance of multi-dimensional structures of the explosion both in theory and observation (e.g., Kasen et al., 2009; Maeda et al., 2010a). Most previous studies also focused on the emission in the MeV range. In this paper, we present our radiation transfer simulations of the high energy emission based on the state-of-the-art SN Ia explosion models. We extend our analysis to hard X-ray and soft $\gamma$-ray regimes, for which dramatic improvement is expected in the observational sensitivities thanks to new generation observatories like {\em NuStar} (Koglin et al., 2005) or {\em Astro-H} (Takahashi et al., 2010). We predict that these telescopes are capable of detecting the radioactive decay signals from SNe Ia, at a rate of once in a year or at least once in a few years. We also briefly comment on perspectives for core-collapse SNe. 

\section{Expected Characteristics}

\begin{figure*}[ht]
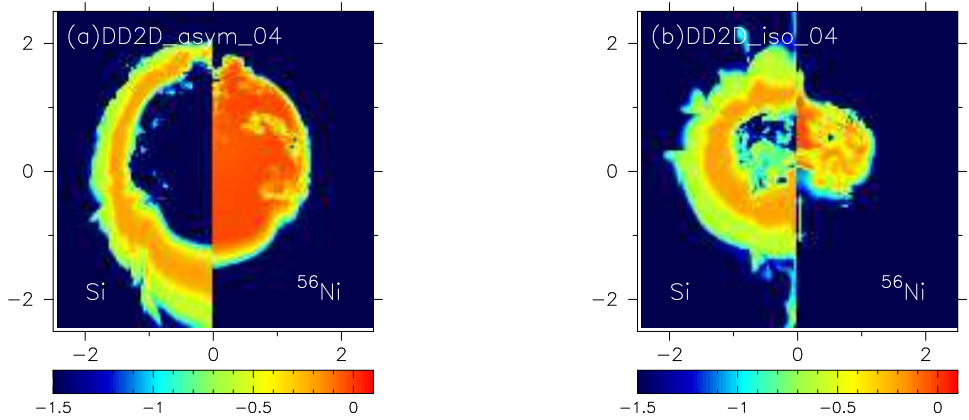

\centerline{
        \begin{minipage}[]{0.45\textwidth}
                \resizebox{50mm}{!}{\includegraphics{maeda_FRAWS_2013_01_fig01a_REVISED.eps}}
        \end{minipage}
       \begin{minipage}[]{0.45\textwidth}
                 \resizebox{50mm}{!}{\includegraphics{maeda_FRAWS_2013_01_fig01b_REVISED.eps}}
        \end{minipage}
}
\caption{Examples of the delayed-detonation models (Kasen et al., 2009; see also Maeda et al., 2010b). The mass fractions of Si (left) and $^{56}$Ni (right) are shown, on a logarithmic scale. The axes are in $10,000$ km s$^{-1}$. }
\label{fig1}
\end{figure*}

We performed radiation transfer simulations (Maeda et al. 2012) based on a series of two-dimensional delayed detonation models by Kasen et al. (2009). The delayed detonation model is among the most popular scenarios for SNe Ia, resulting from a near-center ignition of thermonuclear sparks within a Chandrasekhar-mass white dwarf (Khokhlov, 1991). Conditions for the initial triggers have not been clarified from the first principle (e.g., Seitenzahl et al., 2013), thus Kasen et al. (2009) adopted various conditions (i.e., distribution of the sparks) and produced a range of the ejecta models. In this scenario, different initial conditions can be associated with observed diversities. Figure 1 shows examples of the ejecta structure. The model DD2D\_asym\_04 is for bright SNe Ia (resulting in $\sim 1 M_{\odot}$ of $^{56}$Ni), while DD2D\_iso\_04 is for fainter ones ($\sim 0.4 M_{\odot}$ of $^{56}$Ni). Generally, this model sequence predicts more asymmetric structure for fainter SNe Ia (note that the initial condition of model DD2D\_asym\_04 is indeed more asymmetric than DD2D\_iso\_04, but the post-explosion ejecta are less asymmetric). 

Examples of the synthetic spectra are shown in Figure 2. The spectra are characterized by the decay lines, Compton scattering continuum, and the low energy cut off by the photoelectric absorption. At 20 days, the decay lines from $^{56}$Ni $\to$ $^{56}$Co (e-folding time of $\sim 8.8$ days) are more important than those from $^{56}$Co $\to$ $^{56}$Fe ($\sim 113$ days) as characterized by strong lines in the soft $\gamma$-ray range (e.g., the 158 keV line followed in strength by the 270 and 480 keV lines). Later on, the strong lines are mostly in the MeV range (i.e., the 847 keV line as the strongest) except for the annihilation line. The cut off energy becomes higher as time goes by due to the increasing contribution to the emission from the deeper part where the mean atomic number and photoelectric cross sections are larger. Thus, overall the spectra evolve from soft to hard as time goes by. This indicates that follow-up of SNe at relatively early phases is important in hard X-ray and soft $\gamma$-ray range. The classical 1D model W7 has lower energy cut off than the delayed detonation model sequence, since the W7 model has less extended explosive nucleosynthesis in the surface layer than the delayed detonation model.

Due to increasing transparency, the emission is sensitive to model variants (including the viewing direction) in the earlier phase, while the mass of $^{56}$Ni plays a dominant role in the later phase. Thus, observation at relatively early phases in (relatively) soft bands can provide unique diagnostics in clarifying the explosion mechanism(s). In the multi-D delayed detonation model sequence, a unique prediction is that faint SNe Ia should show larger variations in the high energy emission arising from various viewing directions than brighter ones. Such prediction can be tested once there are at least a few samples of high energy emission detected for SNe Ia. Another diagnostics using the `soft' bands includes the surface composition analysis from the photoelectric absorption, e.g., the different behavior shown for the W7 model and delayed detonation models.

\begin{figure*}[ht]
\centerline{
       \resizebox{140mm}{!}{\includegraphics{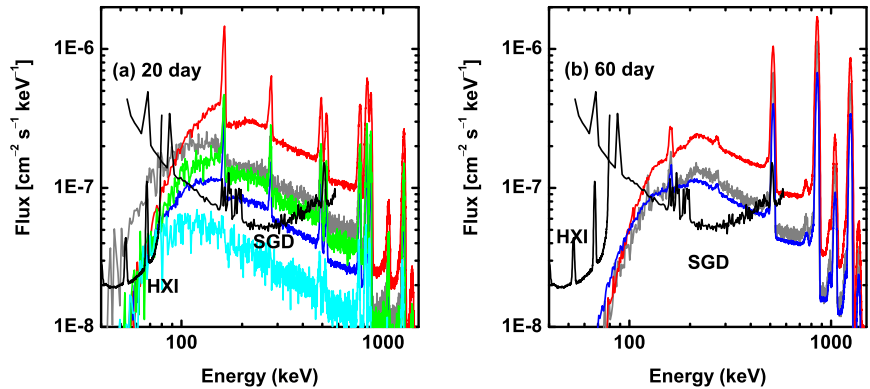}}
}
\vspace{-1.5cm}
\caption{Examples of synthetic spectra at (a) 20 days and (b) 60 days after the explosion.  Shown here are angle-averaged spectra for models DD2D\_asym\_04 (red line), W7 (gray; Nomoto et al., 1984), and DD2D\_iso\_04 (dark blue) (at 10 Mpc).  The angle-dependent spectra seen from two opposite directions are shown for DD2D\_iso\_04 (green and cyan) at 20 days. At 60 days the angle dependence is small. The angle dependence is small for DD2D\_asym\_04 at both epochs. Sensitivity curves for an exposure with $10^6$ seconds of HXI and SGD on board {\em Astro-H} (Tajima et al., 2010; Takahashi et al., 2010) are shown by black lines.
}
\label{fig2}
\end{figure*}

\section{Observational Perspectives}

Table 1 summarizes expected detectability of extragalactic SNe Ia at various band passes by a few current and future instruments. While detecting the MeV lines from the $^{56}$Co decay has been challenged in the past, even with SPI on board {\em INTEGRAL} and $10^6$ s exposure (Roques et al., 2003; see also Isern et al., 2013), this is limited to extremely nearby SNe Ia up to 5 - 6 Mpc (or 8 Mpc for extremely bright SNe Ia). Such nearby events are expected only once in a decade. This frustrating situation in the MeV range will be improved only when the sensitivity is improved by an order of magnitude, hopefully by proposed new generation observatories like {\em GRIPS} (Greiner et al., 2012; see also Summa et al., 2013). 

We propose that new generation hard-X and/or soft $\gamma$-ray instruments can change the situation. In hard X-rays, {\em NuStar} has been already launched. {\em Astro-H} is scheduled for launch in 2015, which will be attached with HXI (hard X-ray) and SGD (soft $\gamma$-ray). These instruments are expected to have $10^6$ s exposure sensitivities sufficient to reach SNe Ia at 15 (conservative estimate) or 25 Mpc (optimistic estimate) (Tab. 1). The `line detection' is more challenging, and we estimate the distance for the $5\sigma$ detection of the 158 keV line is $\sim 10 - 15$ Mpc for Model DD2D\_asym\_04 and $\sim 3 - 5$ Mpc for the other two models shown in Table 1. Figure 3 shows simulations for expected signals from SNe Ia at 15 Mpc, by convolving the synthetic spectra and designed sensitivity curves of HXI and SGD (here the adopted sensitivity curve corresponds to our `optimistic' case). 

In 2011-2012, 6 SNe Ia were discovered within $\sim 20$ Mpc, 3 of which were within $\sim 15$ Mpc (from the Asiago SN Catalog; Barbon et al., 1999). Most of these were discovered soon after the explosion, and especially the nearest ones were all discovered within a week after the explosion -- thus, ToO follow-up at the hard X and soft $\gamma$-ray peak (2 - 3 weeks after the explosion) is feasible. With $10^{6}$ s exposure, we predict that a few (optimistic) or one (conservative) SNe Ia per year are reachable by {\em Astro-H}. 

\section{Discussion and Conclusions} 

\begin{table*} [!ht] 
\begin{center}
\caption{Expected Detectability (for an exposure of $10^6$ s centered at the peak date in each band pass). Shown here are limiting distance and the expected number of SNe Ia within the distance (shown in parenthesis). `cons' and `opt' are conservative and optimistic estimates, respectively. See Maeda et al. (2012) for details.}
\bigskip
\begin{tabular}{lllll}
\hline\hline
          &     &   DD2D\_asym\_04 & W7 & DD2D\_iso\_04\\\hline
          & $M$($^{56}$Ni)/$M_{\odot}$  & 1.02 & 0.64 & 0.42\\\hline
Band (keV) & Instrument & Mpc (SNe year$^{-1}$) & &\\\hline
60--80 & HXI & 13.9 (0.43) & 17.7 (0.96) & 10.5 (0.09)\\
          & NuStar (cons.) & 13.0 (0.43) & 16.5 (0.70) & 9.7 (0.09)\\
          & NuStar (opt.) & 18.4 (1.13) & 23.3 (2.52) & 13.8 (0.43)\\
158 & SPI & 4.6 ($<$0.09) & 2.9 ($<$0.09) & 2.3 ($<$0.09)\\
      & SGD (cons.) & 22.2 (2.09) & 14.2 (0.43) & 11.4 (0.09)\\
      & SGD (opt.) & 38.5 (6.70) & 24.6 (2.96) & 19.7 (1.57)\\
200--460 & SPI & 3.7 ($<$0.09) & 2.7 ($<$0.09) & 2.3 ($<$0.09) \\
             & SGD (cons.) & 11.6 (0.09) & 8.6 (0.09) & 7.1 (0.09) \\
             & SGD (opt.) & 20.2 (1.74) & 14.8 (0.43) & 12.3 (0.26) \\
812 & SPI & 4.3 ($<$0.09) & 2.6 ($<$0.09) & 2.0 ($<$0.09)\\
      & GRIPS & 16.8 (0.87) & 10.0 (0.09) & 7.6 (0.09)\\
847 & SPI & 7.7 (0.09) & 5.4 ($<$0.09) & 4.6 ($<$0.09)\\
      & GRIPS & 29.8 (4.52) & 21.0 (2.00) & 18.0 (1.04)\\\hline
\end{tabular}
\end{center}
\end{table*}

\begin{figure*}[ht]
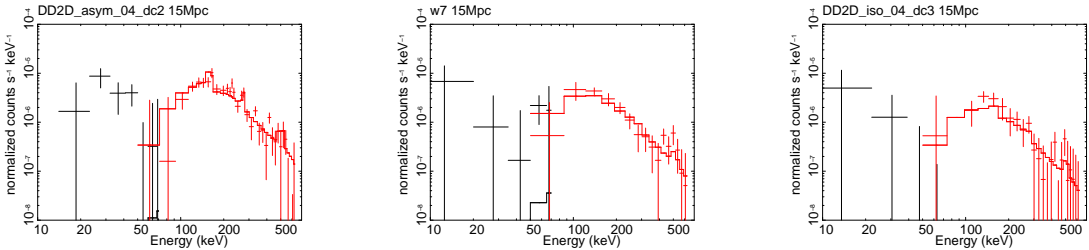

\centerline{
        \begin{minipage}[]{0.3\textwidth}
                \resizebox{30mm}{!}{\includegraphics{maeda_FRAWS_2013_01_fig03a_REVISED.ps}}
        \end{minipage}
       \begin{minipage}[]{0.3\textwidth}
                 \resizebox{30mm}{!}{\includegraphics{maeda_FRAWS_2013_01_fig03b_REVISED.ps}}
        \end{minipage}
        \begin{minipage}[]{0.3\textwidth}
                 \resizebox{30mm}{!}{\includegraphics{maeda_FRAWS_2013_01_fig03c_REVISED.ps}}
        \end{minipage}
}
\caption{Detector response simulations for an exposure of $10^{6}$ seconds for selected models (Tab. 1), for HXI (black) and SGD (red) on board {\em Astro-H}. The model spectra at 20 days after the explosion are used as input, placed at distances of 15 Mpc. The sensitivity curves are adopted from Kokubun et al. (2010), Tajima et al. (2010), and Takahashi et al. (2010). Note that the photon count is very low in the HXI band for all the models at this distance, thus an apparent detection by HXI (left panel) just comes from the statistical fluctuation. 
}
\label{fig1}
\end{figure*}

According to our simulations of radioactive decay signals from SNe Ia, the new generation hard X-ray and soft $\gamma$-ray observatories (either {\em NuStar} or {\em Astro-H}) are expected to be capable of detecting these signals from SNe Ia up to $\sim 20$ Mpc with $10^6$ s exposure. This will hopefully lead to nearly annual detections, dramatically changing the field. We thus propose follow-up of nearby SNe Ia by these telescopes in a ToO mode. With a standard set up with a few $10^{5}$ s exposure, the detection will be limited to extremely nearby objects (i.e., up to $\sim $8 - 12 Mpc for SNe Ia with average brightness), but still there is a good chance of first solid detection of the signal from SNe Ia. Once detected, it will provide various diagnostics on explosive nucleosynthesis and explosion mechanisms, and here a combination of hard X-ray and soft $\gamma$-ray will be essential. 

A similar argument applies for core-collapse SNe. By combining results from similar simulations for core-collapse SNe (Maeda, 2006) and those obtained for SNe Ia (Maeda et al., 2012), we find the following. SNe IIp (an explosion of a red supergiant) is not a promising target in the soft bands, since the thick H envelope is still opaque in the early phase where the $^{56}$Ni decay can provide the strong emission in these band passes. Among different types of core-collapse SNe, SNe IIb/Ib/Ic (an explosion of a He or C+O star) are most promising. We estimate that the peak date in the high energy emission will be similar to (or a bit delayed than) that of SNe Ia (i.e., 2 - 3 weeks since the explosion). The smaller amount of $^{56}$Ni here (typically $\sim 0.1 M_{\odot}$), taking into account the delayed peak date, results in the expected horizon as $5 - 8$ Mpc (conservative and optimistic). Such nearby objects are rare, but expected to be discovered at a rate of a few in a decade according to the past statistics.

\thanks
KM thanks Franco Giovannelli and the organizers of Frascati Workshop 2013 for creating the friendly and stimulating atmosphere. The work by KM has been supported by WPI initiative, MEXT, Japan. The authors acknowledge financial support by Grant-in-Aid for Scientific Research from MEXT (22684012, 23340055, 23740141).

\end{multicols}
\end{document}